\documentstyle[preprint,prabib,aps,12pt]{revtex}
\begin{document}
\title{Multidimensional geometrical model of the renormalized \\ 
electrical charge with splitting off \\
the extra coordinates}
\author{ Dzhunushaliev V.D.
\thanks{E-mail address: dzhun@freenet.bishkek.su}}
\address {Department of the Theoretical Physics \\
          Kyrgyz State National University, Bishkek, 720024}
\maketitle
\begin{abstract}
A geometrical model of electric charge is proposed. This model has 
``naked'' charge screened with a {\em ``fur - coat''} consisting of 
virtual wormholes. The 5D wormhole solution in the Kaluza - 
Klein theory is the ``naked'' charge. The splitting off 
of the 5D dimension happens on the two spheres (null surfaces) 
bounding this 5D 
wormhole. This allows one to sew two  Reissner - Nordstr\"om 
black holes 
onto it on both sides. The virtual wormholes entrap a part 
of the electrical 
flux lines coming into the  ``naked'' charge. 
This effect essentially 
changes the charge visible at infinity so that 
it satisfies the real 
relation $m^2<e^2$.
\end{abstract}

\section{Introduction}
The existence of an intrinsic structure for the electron may be a  
radical remedy for the singularities in quantum field theory. 
In this regard
Wheeler \cite{wh1} - \cite{wh3} proposed a model of 
``charge without 
charge''. The idea was that a wormhole (WH), which 
entraps electrical flux 
lines, appears as an electrical charge for an external observer. 
The sign of
the charge depends upon how the WH part is observed: 
with ingoing or outgoing 
flux lines. The first case represents a negative charge and  
the second case represents a positive charge. 
\par
Wheeler's model is 4 dimensional. On the other hand, 
the 5 dimensional 
Kaluza - Klein theory has the wonderful property 
that the ``electrical'' 
field is a purely geometric object. In this essay we 
will try combine 
these two viewpoints of the electrical field. A $5D$ solution 
of the Kaluza - Klein theory is given in Ref.\cite{dzh1}; 
it is a Lorentzian 
WH, and it is bounded by two null surfaces. 
This WH can be sewn together 
with two  stationary Reissner - Nordstr\"om solutions. 
The resulting object 
is a composite WH with following properties: 
\begin{enumerate}
\item
The interior part is a $5D$ Lorentzian WH.
\item
The two exterior parts are the 4D asymptotical flat regions of a 
stationary Reissner - Nordstr\"om solution (where $r>r_+, r_+ = m + 
\sqrt{m^2 - e^2}$ is the event horizon). 
One region contains the electric 
flux lines which enter the $5D$ WH and the other region contains the  
electric flux lines which exit the $5D$ WH. 
\item
The sewing together of the interior region with each exterior region 
happens on the event horizons $r_+$. This guarantees 
the nonobservability 
of the multidimensional world for exterior observer. 
\item
The $5D$ extra coordinate splits off of the surface sewing. 
This results in the $G_{5t}$ component of the intrinsic metric 
tensor generating the $4D$ electrical field in 
the exterior regions. 
\end{enumerate}
Similar objects were considered in Ref. \cite{gun3}. 
\par 
Currently, it is often suggested that 
an important feature of modern Grand Unified Theories (GUT) are 
their multidimensionality (MD). In these theories our Universe 
is a MD Universe with compactified extra dimensions. MD gravity on 
fibre bundles with the fibres = gauge group have been 
investigated intensively for roughly 80 years. In Ref's 
\cite{per}-\cite{coq} it was established 
that MD gravity is equivalent to 4D gravity + 
gauge field + scalar field. The multidimensionality is 
a very serious component for almost all recent GUT. 
In these theories the superfluous extra dimensions (ED) are 
frozen and contracted to very small size. In this essay we conjecture
that there are some regions with noncompactified ED in 
our MD Universe. This can be true near 
either a pointlike or a cosmology singularity. 
In the first case such domains with noncompactified ED can be found 
under the event horizon. 
\par 
In Ref. \cite{gun1} an object called the Gravitational Bag was 
investigated. In such an object string excitations can be 
enclosed inside the Gravitational Bag. The MD metric 
for the Gravitational 
Bag does not have off-diagonal components, which is the opposite 
of the case investigated here. The presence of off-diagonal 
components of the MD metric (according to theorem stated 
below these components are ``the gauge potential'') leads to the 
conclusion that the MD metric has a wormhole-like appearance, and 
this solution is located between two null surfaces. 
In Ref. \cite{gun2} a 5D model was considered 
in which a bubble of a spacetime with a small compact 
dimension is sewn onto a spacetime with all macroscopic 
dimensions. 
\par
This essay addresses the following question: how can the regions with 
noncompactified ED remain in the turned on state even though 
compactification has taken place in our MD Universe? We will show 
that these are regions with very strong gravitational fields.

\section{The composite wormhole as an electrical charge}

To begin we will review some of the results of Ref.\cite{dzh1}.  
The $5D$ metric has the following WH-like appearance:
\begin{equation}
ds^2 = e^{2\nu (r)}dt^2 - e^{2\psi (r)} \left (d\chi - \omega (r)dt
\right )^2 - dr^2 - a^2(r)\left (d\theta ^2 + 
\sin ^2\theta d\varphi ^2
\right ),
\label{11}
\end{equation}
where $\chi$ is extra  $5^{th}$ coordinate; $r, \theta ,\varphi$ 
are 3D polar coordinates; $t$ is time. The corresponding 
$5D$ Einstein's  
equations have the following solution:
\begin{eqnarray}
a^2 & = & r^2_0 + r^2,
\label{12a}\\
e^{-2\psi}= e^{2\nu} & = & {{2r_0}\over q} {{r^2_0 + r^2}
\over{r^2_0 - r^2}},
\label{12b}\\
\omega &= & {{4r^2_0}\over q} {r\over{r^2_0 - r^2}},
\label{12c}
\end{eqnarray}
where $r_0$ is the throat of the wormhole; $q$ is a 5D 
``electrical'' charge. It is easy to see that the time component of 
metric tensor $G_{tt} (r=\pm r_0)=0$. This indicates that this is a 
null surface, since on it $ds^2=0$. The sewing together 
of the $5D$ and 
$4D$ physical quantities is accomplished in the following manner:
\begin{eqnarray}
e^{2\nu _0} - \omega ^2_0 e^{-2\nu _0} = G_{tt}\left 
(\pm r_0\right ) = g_{tt}\left (r_+\right ) = 0,
\label{13a}\\
r^2_0 = G_{\theta\theta}(\pm r_0) = g_{\theta\theta}(r_+) = r^2_+,
\label{13b}
\end{eqnarray}
where $G$ and $g$ are $5D$ and $4D$ metric tensors respectively, and 
$r_+$ is event horizon for the Reissner - Nordstr\"om 
solution. The quantities with the $(0)$ subscript are taken to be
evaluated at $r=\pm r_0$. 
\par
To connect $G_{\chi t}$ with the 4D electric field we consider the 5D 
$R_{\chi t}=0$ equation and Maxwell's equation:
\begin{eqnarray}
\left [a^2\left (\omega 'e^{-4\nu}\right )\right ]' = 0,
\label{23}\\
\left (r^2E_r\right )' = 0,
\label{24}
\end{eqnarray}
here $E_r$ is the 4D electrical field, and the primes indicate 
differentiation with respect to $r$.
\par
These two equations are essentially Gauss's law, and they indicate that
some quantity multipled by area is conserved. In the 4D case this quantity 
is the 4D Maxwell's electrical field, and from this it follows that the 
electrical charge is conserved. Thus, we can naturally 
join the 4D electrical Reissner - Nordstr\"om 
field, $E_{RN} = e/r^2_+$, with ``electrical" Kaluza - Klein field, 
$E_{KK} = \omega 'e^{-4\nu}$, on the event horizon:
\begin{equation}
\omega _0'e^{-4\nu _0} = {q\over{2r^2_0}} = E_{KK} = 
E_{RN} = {e\over{r^2_+}}.
\label{25}
\end{equation}
The Reissner - Nordstr\"om condition $m^2>e^2$ is one of the basic 
causes obstructing the interpretation of such a composite WH as a model
for the  electron. 
\par
Quantum gravity indicates that at distances close 
to the Planck length the 
metric fluctuations are so large that the topological 
fluctuations - wormholes 
(handles) -  appear in spacetime. Such fluctuations spring up in 
regions with very strong gravitational field, such as near  
the event horizon surface $r=\pm r_+$ of a sufficiently 
small black hole. 
\par
We propose a model which assumes that virtual wormholes (VWH) 
arise between two 4D regions with incoming and outcoming 
flux lines, and 
which has a strong gravitational field near the two surfaces 
$r=\pm r_+=\pm r_0$. The appearance of these VWH indicates that they 
entrap part of the $4D$ electrical flux lines. 
If the total cross sectional 
size of all VWHs is of the same order as the cross 
sectional size of the 
$5D$ WH, then the VWHs can entrap almost all 
electrical flux lines coming into 
the $5D$ WH. In this case the exterior observer 
can detect electrical 
charge with the real relation between mass and charge 
of $m^2 < e^2$.
\par
Thus, in this model the real 
electrical charge consists of a ``naked'' electrical 
charge ($5D$ WH) dressed
with a {\em ``fur - coat''} of VWHs. Such a {\em ``fur - coat''} 
essentially changes the electrical charge visible 
at infinity, i.e. it can 
be viewed as the quantum renormalization of the charge. 
This model cannot 
yet describe the charge quantization and $\hbar /2$ 
spin of the electron. In 
the first case it is necessary to have a quantum field theory of 
gravity which we do not yet have. The second case 
is discussed below.

\section{Compactification mechanism}

In this section the compactification of the above 
model is proposed. 
This idea is based on the suggestion that the our Universe 
is a multidimensional Universe with compactified extra 
dimensions (i.e. the extra dimensions are nondynamical variables of
the full multidimensional metric).  
However, there can be domains where the extra dimensions are 
noncompactified and the metric is dynamical. The compactification 
takes place on the boundary of these domains. 
Such a Universe is a principal bundle 
with the gauge group as the fibre. The underlying idea suggested 
here is the following: {\bf The linear 
sizes of the fibre are frozen almost everywhere but there are 
domains where the linear sizes of the fibre depends on the  
point of the base.} 
\par 
We now present the following theorem which will be important 
for us \cite{per}-\cite{coq}: 
\par
Let the group $G$ be the fibre of a principal  bundle.  Then  
there  is  a  one-to-one 
correspondence between $G$-invariant metrics on the  total  space
${\cal X}$ 
and the triplets $(g_{\mu \nu }, A^{a}_{\mu }, h\gamma _{ab})$, 
where $g_{\mu \nu }$ is Einstein's pseudo  - 
Riemannian metric, $A^{a}_{\mu }$ is a gauge field  of  
the $G$  group, 
and $h\gamma _{ab}$ a symmetric metric on the fibre. In this case 
we can write down the MD Ricci scalar $R^{(MD)}$ in the following 
way:
\begin{equation}
 R^{(MD)} = R^{(4)} + R^{(G)}  -
 \frac{1}{4}F^a_{\mu\nu}F^{a\mu\nu}
 - d \partial _\mu (g^{\mu\nu}\varphi^{-1}\partial _\nu\varphi )-
  \frac{d(d+1)}{4h^2}\partial _\mu \varphi \partial ^\mu \varphi ,
\label{1-1}
\end{equation}
where $R^{(4)}$ is the Ricci scalar of Einstein's 4D spacetime; 
$R^{(G)}$ is the Ricci scalar of the gauge group $G$, 
$F^a_{\mu\nu} = \partial _\mu A^\nu - \partial _\nu A^a_\mu -
f_{abc}A^b_\mu A^c_\nu $ is the gauge field strength, 
$d$ is the dimension of 
the gauge group, $\nabla _\mu$ is the covariant derivative 
on the 4D spacetime, $f_{abc}$ are the structure constants of the 
gauge group, $\varphi$ is the linear size of the 
fibre and the Ricci 
tensor $R_{\mu\nu} = R^\alpha_{\mu\nu\alpha}$. Greek indices 
are spacetime indices and Latin are indices are on the fibre. 
\par 
This theorem indicates that the domains with compactified 
and noncompactified EDs differ in that the scalar function 
$\varphi (x^\mu)$ is constant in the first case and is a 
dynamical function in the second case. 
In this second case the off-diagonal components of the MD metric 
can be continued to the 4D region as physical gauge fields 
(electromagnetic, SU(2) or SU(3)) in the following way: 
we can attach two black holes to the two null surfaces $\pm r_0$ 
of the  wormhole-like solutions, 
base to base and fibres to fibres. Of course we must have a 
compactification mechanism on the boundary between domains 
with compactified and noncompactified ED. 
\par 
What is the cause of such compactification? We presume that: 
{\bf in a strong gravitational field 
the metric on the extra dimensions is a dynamical 
variable and by decreasing 
the gravitational field it becomes nondynamical.} 
It is very important to underscore that this process is 
a completely quantum jump not a classical step-by-step 
process! The qualitative description of such a phenomena 
can be described in the following way \cite{dzh4}: 
\par 
Every physical object in our Universe can be described as 
some program.  For  example,  the  Universe  in  Einstein's
theory is  characterized  by  an algorithm  defining  all
notions that lie at the base of  the  Universe.  In  this  case 
the topological  space, its  geometrical  structures   (metrics,
connection), Einstein's equations and so on must be described
by an algorithm. Such an algorithm may be realized with some 
universal machine (e.g. a Turing machine). 
The length of such a minimal 
algorithm is called the algorithmical complexity of the given object.
\par
In the 1960s  A.N. Kolmogorov investigated the notion 
of probability arising 
from this algorithmic  viewpoint. The classical definition 
of probability is  
connected  with the calculation of a ratio of the number 
of ways in which the 
trial can succeed to the total number of outcomes of 
the trial. Kolmogorov 
investigated  the   notion   of probability from this 
algorithmic viewpoint. 
He  wrote  in\cite{kol}:
\begin{enumerate}
\item
the basic notions of the information theory must and  may
be defined without the application of probability theory, so that the 
``entropy'' and ``number of information'' notions are  applicable
to the \underline{individual objects};
\item
the notions of information theory which are introduced in this
manner can form the basis  of a new  conception of  
chance, corresponding to the natural idea that chance is the absence 
of laws.
\end{enumerate}
\par
Kolmogorov's idea can be of interest in quantum gravity since 
the notion of probability is adaptable to a single object. From 
Kolmogorov's viewpoint {\em ``chance'' = ``complexity''}, 
and  therefore  
the probability in quantum gravity can be  connected  with
the complexity of a given structure. The exact definition 
of algorithmic  
complexity given by Kolmogorov is \cite{kol}:
{\em
\par
The algorithmic complexity $K(x\mid y)$ of the  object $x$  for a 
given object $y$ is the minimal length of the ``program'' $P$
that is written as a sequence of  the  zeros  and  unities
which allows us to construct $x$ having $y$:
\begin{equation}
K(x\mid y) = \min_{A( P,y)=x} l(P)
\label{11a}
\end{equation}
where $l(P)$ is length of the  program $P$; $A(P,y)$  is  the
algorithm  for calculating an object $x$, using  the  program $P$,
when the object $y$ is given.
} 
\par 
Furthermore, we  shall  assume that the algorithmic complexity of 
a given object can be spontaneously changed. In this 
case the situation  is analogous to the spontaneous 
electron transition 
between  two energy levels in an atom: an electron that was in the  
exited-state level $E_{1}$ (quantum gravitational object with the 
algorithmic complexity $K_{1}$) falls,
as a result of the spontaneous quantum transition, 
to  the ground-state  
level $E_{2}$ (appears as a  quantum  gravitational object 
with lesser 
algorithmic complexity $K_{2}< K_{1}$). It is probable 
that the splitting  off 
of  the extra dimensions is such a quantum transition 
with a changing 
algorithmical complexity.

\section{Discussion}

There are 2 basic problems related with the above constructed WHs with 
electrical (colour) charge:
\begin{enumerate}
\item
What fraction of the flux lines which arrive to the 
$5D$ WH are entrapped 
by the VWH {\em ``fur - coat''}?
\item
What is the spin of such a composite wormhole?
\end{enumerate}
The first question can only be solved within the framework of a full 
theory of quantum gravity. Wheeler has underscored repeatedly the 
importance of 
a geometrical model of spin $\hbar /2$. He wrote that the geometrical 
description of $\hbar /2$ spin must be a significant component 
of any electron model! It is possible that one may need 
to include anti-commuting coordinates in the spacetime,
i.e. the appearance of $\hbar /2$ spin may be connected with the 
existence odd anti-commuting coordinates in our Universe. 

\section{Acknowledgments}
This project has been funded in part by the National 
Council under the Collaboration in Basic Science 
and Engineering Program 
and in part by ISTC Grant KR-154.

\newpage

\section{The information concerning author}

Home address:	mcr.Asanbai, d.25, kv.24, Bishkek, 720060, 
Kyrgyz Republic.
\par 
Temporary address: 

\end{document}